\documentclass{aa501}
\usepackage{graphics}
\usepackage{graphicx}
\usepackage{amssymb}
\begin{document}

\def\msyr{M$_{\odot}~$yr$^{-1}$}
\def\ha{H$_{\alpha}$~}
\def\hh{${\rm H}_{2}$~}
\def\ico{$I_{\rm CO}$~}
\def\b14{$B_{1.4}$}
\def\ssfr{$\Sigma_{\rm SFR}$~}
\def\shh{$\Sigma_{\rm H_{2}}$~}
\def\sfe{${\rm SFE_{\tau_{8}}}$}

\title{Radio continuum and CO emission in star-forming galaxies.}

\author{M. Murgia\inst{1}
\and A. Crapsi\inst{1,2}
\and L. Moscadelli~\inst{3}
\and L. Gregorini \inst{1,4}
}

\institute{Istituto di Radioastronomia del CNR, Via Gobetti 101, I-40129, 
Bologna, Italy
\and Osservatorio Astrofisico di Arcetri, Largo E. Fermi 5, I-50125, Firenze,
Italy
\and Osservatorio Astronomico di Cagliari, Loc. Poggio dei Pini, Strada 54, 
I-09012 Capoterra (CA), Italy 
\and Dipartimento di Fisica, Universit{\`a} di Bologna, Via B. Pichat 6/2,
 I--40127 Bologna, Italy
}

\date{Received; Accepted}

\abstract{
We combine the radio continuum images from the NRAO VLA Sky Survey with the 
CO-line observations from the extragalactic CO survey of the Five College Radio
 Astronomy Observatory to study the relationship between molecular gas and 
the star formation rate within the disks of 180 spiral galaxies
 at 45\arcsec~resolution. We find a tight correlation between 
these quantities. On average, the ratio between the radio continuum and 
the CO emission is constant, within a factor of 3, both inside the same
 galaxy and from galaxy to galaxy. The mean star formation efficiency deduced 
from the radio continuum corresponds to convert 3.5\% of the available
 molecular gas into stars on a time scale of $10^{8}$ yr and 
depends weakly on general galaxy properties, such as Hubble type or 
nuclear activity. A comparison is made with another similar analysis
 performed using the \ha luminosity as star formation indicator. The overall
 agreement we find between the two studies reinforces the use of the radio
 luminosity as star formation rate indicator not only on global but also on
 local scales.
}

\offprints{M. Murgia, \email{murgia@ira.bo.cnr.it}}

\titlerunning{Radio continuum and CO emission in star-forming galaxies.}
\authorrunning{M. Murgia et al.}
\maketitle

\section{Introduction}
One of the major goals of the studies of external galaxies is understanding 
the relationship between the star formation rate (SFR) and the physical condition 
in the interstellar medium.
Several indicators have been suggested to estimate the SFR (of massive stars) 
in galaxies. These include the U-band magnitude, the strength of Balmer lines  
emission, the far-infrared (FIR) emission and the radio luminosity; the rates
 inferred from the different indicators span almost four orders of magnitude
 going from $10^{-2}$ to $10^{2}$ \msyr. Cram 
et al. (1998) checked  the consistency between the SFR deduced from the 
 U-band, H$_{\alpha}$, FIR and radio luminosity using a sample of 700 local 
galaxies. They noted that there are systematic differences
 between these various indicators. In particular they suggested that
 the H$_{\alpha}$ luminosity may underestimate the SFR by  approximately
 an order of magnitude when the SFR is $\geq 20$ \msyr.
They concluded that the radio continuum luminosity at decimeter wavelengths
 of a star forming galaxy provides a better way to estimate the current rate
 of star formation. The radio continuum emission at 1.4 GHz from a star-forming
galaxy is mainly synchrotron radiation produced from relativistic electrons
 accelerated by supernovae explosions (Lequeux 1971). Indeed the radio 
continuum luminosity appears to be directly proportional to the rate of formation 
of supernovae (Condon 1992). This view is reinforced by the tight correlation existing
 between the radio luminosity and the FIR for spiral galaxies 
(see e.g. Condon 1992). Since the radio continuum at 1.4 GHz does not 
suffer significant extinction, the radio luminosity constitutes a very useful
 tool to determine the current SFR in a spiral galaxy. 

Since the discovery that stars form in molecular clouds, it is essential
to determine, not only the rate, but also the efficiency of conversion of the 
interstellar gas in stars; i.e. the star formation efficiency 
(SFE). The SFE measures the formation rate of young stars per unit of mass of gas
 available to form those stars. 
Determining the SFE is important to distinguish a situation in which 
a high SFR indicates a higher efficiency in converting 
 gas in stars rather than a higher gas quantity.

The CO molecule luminosity  and the virial mass of giant molecular clouds
 correlate very well in our Galaxy and in other nearby spirals 
(Young \& Scoville 1991 and references therein).
 The comparison of different SFR tracers with the mass of
 molecular clouds provides indeed an important tool to investigate
 the behaviour of the SFE within and among galaxies.
 Many studies have been concerned with 
the behaviour of the star formation process on global scales, averaged
 over the entire star-forming disk. These works showed that the disk-averaged 
star formation process is well described by a Schmidt (1959) law of the type 
$\Sigma_{\rm SFR} \propto  \Sigma_{\rm gas}^{N}$, where 
$\Sigma_{\rm SFR}$ and $\Sigma_{\rm gas}$ are the observable surface
 density of SFR and total (atomic + molecular) gas density, respectively, 
 and the exponent $N$ typically ranges from 1.3 to 1.5 (Kennicutt 1998).

An interesting development of these global studies, the investigation
 of the behaviour of the SFE {\it within} the disks of the individual
 galaxies, provides much physical insight into the star formation process 
itself. The extragalactic CO survey of the Five College Radio Astronomy 
Observatory (Young et al. 1995, hereafter FCRAO CO Survey) provided a uniform
 database of CO data for 300 galaxies at a resolution of 45\arcsec, opening the
 possibility to extend the study of the Schmidt relationship of the
 SFR versus the H$_{2}$ density over the same physical regions well inside
 the galaxy disks. Since the star formation process involves the 
molecular gas directly, some authors recognized that the determination 
of the Schmidt law assumes a clear physical meaning if restricted to this 
gas component. Moreover, in the considered regions the molecular gas is 
dominant over the atomic one and, contrary to this latter, its azimuthally
averaged distribution follows closely the radial profiles of the main SFR 
indicators (Tacconi and Young 1986, Young \& Scoville 1991).
Rownd \& Young (1999; hereafter RY99) conducted an H$_{\alpha}$ imaging
 of 121 of these galaxies, determining the local relationship between
 the SFR and the molecular gas. They found a correlation between these two 
quantities and concluded that for face-on spiral, in
 general, there are no strong SFE gradients across the star-forming disks. 
The majority of large SFE variations they found are seen between adjacent 
disk points, reflecting regional differences in the SFE,
 and any radial gradients are at most a secondary effect.
In contrast, they pointed out that consistent radial variations 
(up to an order of magnitude or more) of the SFE exist within many highly 
inclined galaxy disks. They attributed the decreasing SFE towards the centers
 of these galaxies to a large amount of dust extinction on the \ha luminosity.

Adler et al. (1991) found a correlation between the radio continuum flux
 density at 20 cm and the CO line emission on global scales for a sample
 of 31 spiral galaxies. 
They also studied the relationship of these two quantities within the disks of
 8 nearby well resolved spiral galaxies, finding that their ratio is 
constant both inside the same galaxy and from galaxy to galaxy.\\

The work we present here is complementary to the analysis of RY99 and extends
 that of Adler et al. (1991).
 We combined the radio continuum images at 1.4 GHz from the NRAO 
VLA Sky Survey (NVSS, Condon et al. 1998) with the FCRAO CO survey to study 
the relationship between the radio continuum and the molecular gas
 point-to-point within the disks of 180 star-forming spiral galaxies.
It is important to stress that we are comparing two homogeneous data set 
 with the same angular resolution of 45\arcsec.\\

The paper is organized as follows: in Sect.~2 and Sect.~3 we present the 
sample used and we describe the data analysis, respectively. In Sect.~4 we
 present the results of the statistical analysis and in Sect.~5 we discuss the
 results obtained.\\

We use a Hubble constant H$_0$=50~km~s$^{-1}$Mpc$^{-1}$ throughout the paper.

\section{Sample selection}
To investigate the star formation law within the disks of normal galaxies,
 we combined the data from two public surveys. We use the NVSS for the
 radio continuum and the FCRAO CO survey for the molecular gas,
 respectively.

The NVSS  was performed at 1.4 GHz with the Very Large Array (VLA)  
in D configuration. It has an angular resolution of 45\arcsec~(FWHM),
a noise level of 0.45 mJy/beam (1$\sigma$) and covers all the sky north of 
declination --40\degr.  The shortest baseline is 35 m, corresponding
 to $\simeq$167$\lambda$, therefore structures up to about 10\arcmin~in 
angular size are properly imaged.

The FCRAO CO survey comprises 300 galaxies observed along the major axis
 of the disk for a total of 1412 locations. 
 Most of the galaxies in the survey are spirals or
 irregulars north of declination --25\degr.
 At the frequency of the CO $J=1-0$ transition (115.27 GHz) the FWHM
 of the 14-m FCRAO telescope is 45\arcsec. The weakest line detected depends
 on the width of the line, and hence on the velocity field within the beam.
 The uncertainties on the individual line intensity vary from galaxy to galaxy.
 A conservative
 estimate of the rms noise, including the calibration, baseline removal, 
and the rms noise per channel is about 25\% (the median signal-to-noise ratio
 is 4). 

We note that the two surveys have uniform sensitivity and identical
 angular resolution. This fact circumvents the difficulties deriving from the
 comparison of data from multiple instruments or studies which are subtly
 incompatible either because of inconsistent signal-to-noise ratios or
 unmatched resolution.

The original FCRAO CO survey includes 300 galaxies selected from the RC2 (de
 Vaucouleurs et al. 1976) or the {\it IRAS} database  satisfying at least
 one of the following criteria: i) $B^{0}_{\rm T} < 13.0$, ii) $S_{60} > 5$ Jy
 or iii) $S_{100} > 10$ Jy. Although the FCRAO CO survey is not a complete 
sample in terms of flux-density or volume limit, the observed galaxies cover a
 wide range of luminosity, morphology and environments. For this reason 
 they represent an ideal database to study the behaviour of the star formation
 process and the molecular gas in a wide variety of conditions. 

Since our interest was primarily to investigate the behaviour of star 
formation within the galaxy disks, we have selected, from the
 FCRAO CO survey, a sub-sample of 180 objects for which there were at least
 three different observations of the CO line in the disk.

\begin{figure*}
\begin{center}
\includegraphics[angle=0, width=18cm]{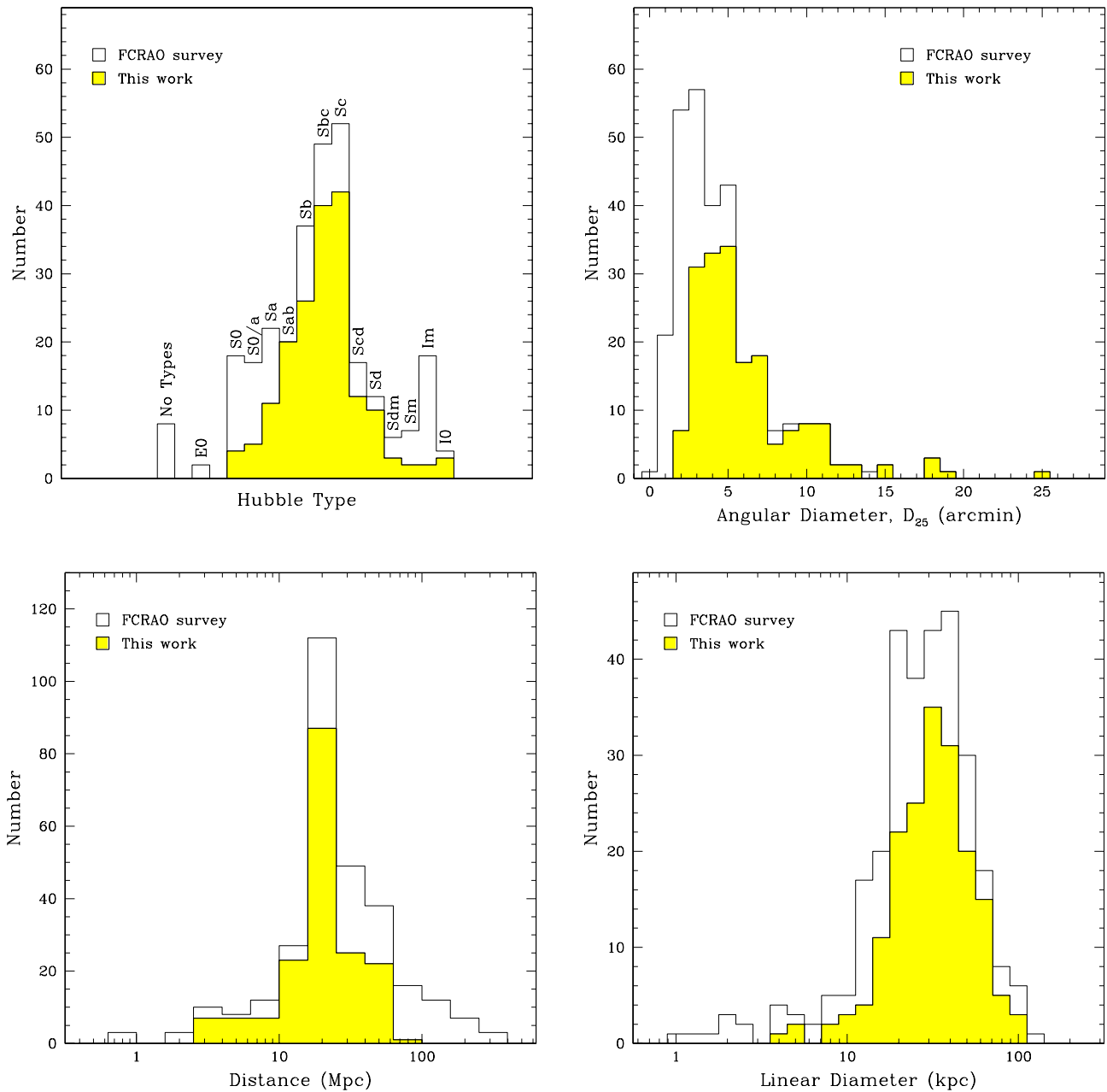}
\end{center}
\caption[]{Distributions of morphological types (top left panel), optical 
angular diameters (top right panel), distances (bottom left panel) and 
linear diameters (bottom right panel): the solid portion of the histograms
 indicates the sub-sample 
used in this work (180 galaxies) whilst the empty portion indicates the 
distribution of the FCRAO CO survey (300 galaxies). Because of its 
 outstanding angular size, the galaxy NGC598 (M33, $D_{25} = 62\arcmin$)
 is not included in the histogram of angular diameters. }
\label{fig1}
\end{figure*}
Distributions of morphological types, angular diameters, distances and 
linear diameters of the sample along with
 the corresponding distributions for the FCRAO CO survey are shown 
in Fig.~\ref{fig1}.

The galaxies in our sample have morphological types ranging between S0 and I0. 
Most of them (89\%) are spiral galaxies with morphological type between Sa-Sd.

The majority of the galaxies (about 90\%) 
have an optical angular diameter (from RC2) $D_{25}\leq$10\arcmin, the median 
angular diameter being about 5\arcmin. This ensures that for all of them the NVSS
 properly recovered the flux density of the extended structures (see above).
The remaining 20 galaxies have an optical angular diameter from 10\arcmin~to
 25\arcmin. For these galaxies it is possible that the NVSS missed a significant 
fraction of flux from the extended structure. Because of its outstanding
angular size, NGC598 (M33, $D_{25} = 62\arcmin$) had been excluded from the
 analysis.

The distances of the galaxies in our sample (taken from Young et al. 1995)
 span from the Local Group up to about 80 Mpc. Over 87 galaxies are at 
 the distance of the Virgo Cluster (20 Mpc).

The linear diameters range from $\sim 4$ to $\sim 100$ kpc. The median
 value is 31 kpc.

 Our selection excluded 
 most of the galaxies with an angular diameter less than 3\arcmin~,
 i.e. the intrinsically small galaxies 
(linear diameter smaller than 30 kpc) and the more distant ones.

The complete list of the galaxies in our study (including the 
galaxy name, Hubble type, inclination, angular diameter and distance) is
 available in electronic format at http://www.ira.bo.cnr.it/
$\sim$crapsi{\_}s/RADIOCO/Article/tab\_art.txt.

\section{Data analysis}
For each galaxy in our sample we extracted the radio surface 
brightness, $B_{1.4}$, from 
the NVSS images . We averaged the NVSS pixel values within circular 
regions of 45\arcsec~diameter, centered on the same pointing positions 
 of the FCRAO survey.
 Following the same formalism
 adopted by RY99 for the \ha emission, the radio brightness is
 expressed in terms of the ratio between the radio luminosity at 1.4~GHz,
 $L_{1.4}$, and the de-projected surface, $\Sigma$, intercepted by the beam
 on the galaxy disk: 
\begin{equation}
\frac{L_{1.4}~({\rm W~Hz^{-1}})}{\Sigma~(\rm kpc^{2})}= 3.2 
\times 10^{18} ~B_{1.4}\left(\frac{\rm mJy}{\rm beam}\right)~\cos~i
\label{lsigmab}
\end{equation}
where $i$ is  the inclination angle of the galaxy with respect to the
 plane of the sky.
The surface luminosity  has the advantage to be distance independent
as opposed to the global luminosity. 

Since most of continuum radio luminosity of normal galaxies is produced
 by relativistic electrons accelerated by supernovae explosion and the 
 supernova rate ($\nu_{\rm SN}$) is directly related to the  SFR of massive 
stars, a relation is expected between the radio 
luminosity and the star formation rate. In the following we indicate with SFR 
the formation rate of stars with mass $M\geq 5 M_{\odot}$.
Condon (1992) calibrated empirically the $\nu_{\rm SN}$-SFR relation using
 the supernova rate and the radio luminosity of our Galaxy:
\begin{equation}
L_{\nu}~({\rm W~Hz^{-1}})=5.3\times10^{21}\nu_{\rm GHz}^{-\alpha}
\cdot{\rm SFR} ~({\rm M_{\odot}yr^{-1}})
\label{condon92}
\end{equation}
where a Miller-Scalo (1979) initial mass function (IMF) is assumed
 and $\alpha$ is the radio spectral index ($S_{\nu}\propto \nu^{-\alpha}$).
Condon (1992) noted that this relation should apply to the majority of
 normal galaxies in order to be consistent with the tight FIR/radio
 correlation widespreadly observed. 

Assuming for the radio spectral index a typical value of $\alpha = 0.8$, from 
Eq.~(\ref{lsigmab}) and Eq.~(\ref{condon92}) the relation between the 
star formation rate per unit surface, $\Sigma_{\rm SFR}$, and the radio
 brightness at 1.4~GHz is found to be:
\begin{equation}
\Sigma_{\rm SFR}\left({\rm \frac{M_{\odot}}{yr~kpc^{2}}}\right)=
8.0\times10^{-4}~B_{1.4}\left(\frac{\rm mJy}{\rm beam}\right)~\cos~i
\label{sfrb}
\end{equation}

In order to derive the \hh~surface density from  the FCRAO CO survey 
 integrated CO intensity \ico , we used the formula (RY99):

\begin{equation}
\Sigma_{{\rm H_{2}}}({\rm M_{\odot}~pc^{-2}})=8.5~I_{\rm CO}
({\rm K~km~s^{-1}})~\cos~i
\label{sigmah2}
\end{equation}
In this formula it is assumed a constant linear conversion factor 
 between the CO luminosity and \hh~mass equal to $X_{\rm CO}=2.8\times 
10^{20}~{\rm ~H_{2}~cm^{-2}[K~km~s^{-1}]^{-1}}$ (Bloemen et al. 1986).\\

A realistic estimate of the uncertainties 
in both $\Sigma_{\rm SFR}$ and $\Sigma_{{\rm H_{2}}}$
 surface densities should consider several systematic effects.
 The relation between the SFR and the radio luminosity is based on many not 
well proved assumptions, such as the IMF thresholds
 and slope and the extrapolation of the $\nu_{\rm SN}$-SFR Milky Way relation
 to other galaxies. Cram et al. (1999) pointed out that different modelling of these
 parameters introduce scaling uncertainties up to a factor of 2.    
 The dominant errors in the gas density are the variation on the
 CO/\hh conversion factor. These variations can be as high as $\pm$40\% for
 luminous spiral galaxies as those studied in this work 
(Devereux \& Young 1991).
Despite these uncertainties, the data provide very strong constrains on
 the form of the SFE because of the wide ranges of SFR and 
 gas densities explored. 

Consistently with the definition given by RY99, the radio SFE is
 calculated by the ratio of \ssfr to \shh. In terms of our observables 
($B_{1.4}$ and \ico) the SFE is expressed by
\begin{equation}
{\rm SFE~(yr^{-1})}\equiv \frac{\Sigma_{\rm SFR}}{\Sigma_{\rm H_{2}}}=
9.4\times 10^{-11} ~\frac{B_{1.4}}{I_{\rm CO}} 
\label{SFE}
\end{equation}
where $B_{1.4}$ and \ico are measured in mJy/beam and K~km~s$^{-1}$,
 respectively.

The SFE defined by Eq.~(\ref{SFE}) gives the fraction of molecular 
gas converted to massive stars per year. Since the typical lifetime
 of the synchrotron radiating electrons is shorter than $10^{8}$~yr 
(Condon 1992), 
  the SFR inferred from the radio luminosity traces a stellar
 population not older than this timescale (hereafter $\tau_{8}$).
 The percentage of molecular gas consumed over all this period is
\begin{equation}
{\rm SFE_{\tau_{8}} (\%)}=0.94 ~\frac{B_{1.4}}{I_{\rm CO}} 
\label{SFEperc}
\end{equation}

\begin{figure*}
\begin{center}
\includegraphics[angle=0, width=17.5cm]{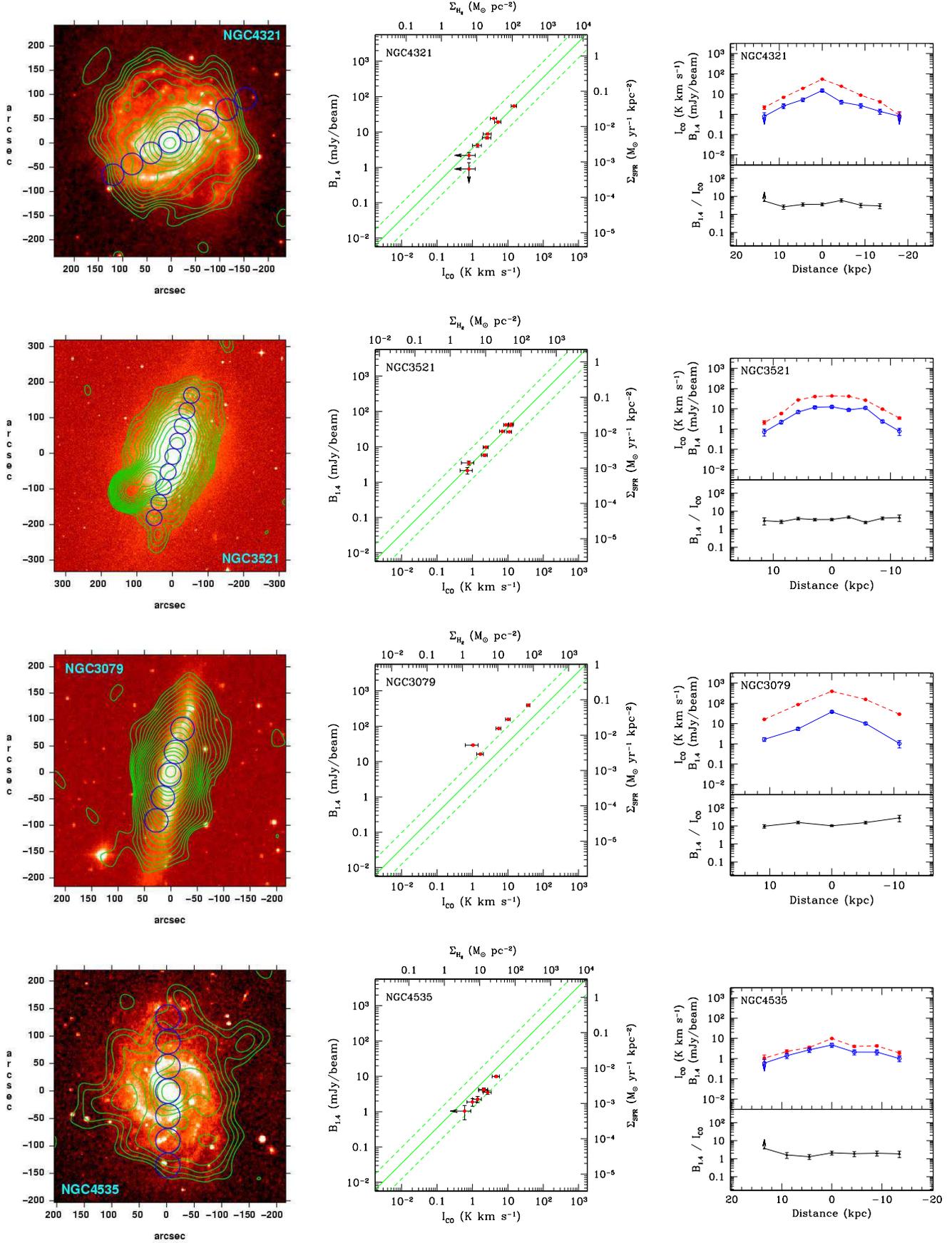}
\end{center}
\caption[]{Examples of selected galaxies, see text. Solid and short-dashed
 lines in middle panels refer to the mean and standard deviation of the \sfe~of the entire 
sample. Solid and dashed lines in right panels refer to \ico an \b14 profiles, 
respectively.}
\label{fig2}
\end{figure*}
\begin{figure*}
\begin{center}
\includegraphics[angle=0, width=17.5cm]{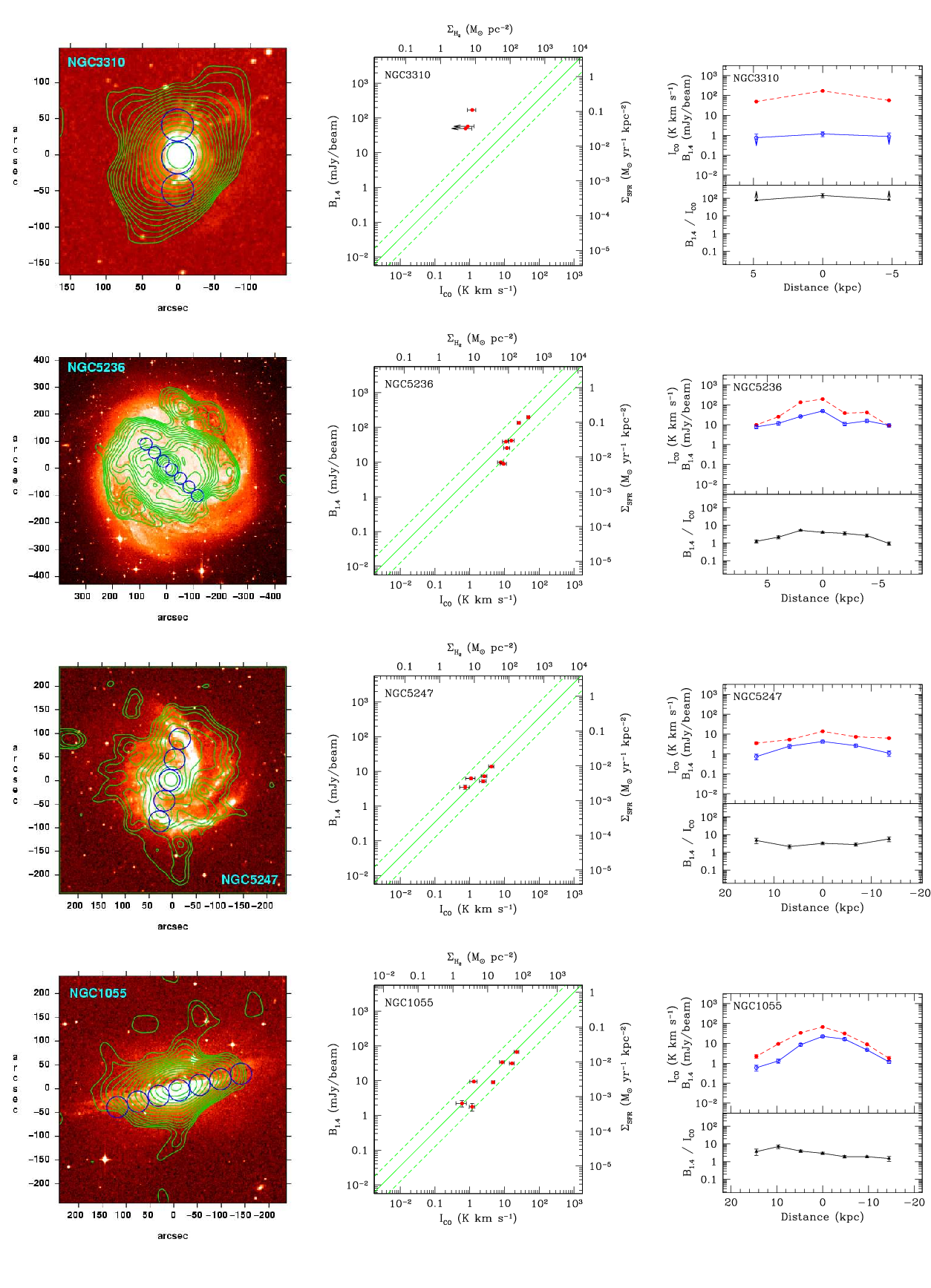}
\end{center}
\caption[]{Examples of selected galaxies, see text. Solid and short-dashed
 lines in middle panels refer to the mean and standard deviation
 of the \sfe~of the entire sample. Solid and dashed lines in right panels refer to \ico an \b14 profiles, 
respectively.}
\label{fig3}
\end{figure*}

\section{Results}
\label{results}
The data analysis presented in Sec.~3 allows one to compare
 the SFR and the molecular gas content to determine the star
 formation efficiency along the galaxy disks. 

Overlays of optical (grayscale) and radio continuum (contours) images for
 eight selected galaxies\footnote{The plots for all the galaxies in the sample 
can be found at the web page:\\ 
http://www.ira.bo.cnr.it/$\sim$crapsi{\_}s/RADIOCO/Catalog} are 
shown in the left columns of Figs. \ref{fig2} and \ref{fig3},
 where the circles indicate
the positions and the beam size of the CO observation.
 In the middle columns we plot $B_{1.4}$ versus \ico reporting also the
 corresponding values of $\Sigma_{\rm SFR}$ and $\Sigma_{{\rm H_{2}}}$ in the
 upper and right axis, respectively.
The panels in the right columns show the radio brightness ($B_{1.4}$), the 
CO integrated intensity (\ico) and their ratio as a function of distance from
 the galaxy center. The convention is that
 radius is positive for positive right ascension (or declination) pointing 
shifts.
The optical images are taken from the red Palomar Digitized Sky Survey. 
The NVSS radio contours start at 0.9 mJy beam$^{-1}$ ($2\sigma$) and are
 spaced by a factor of $\sqrt{2}$. In the plots, error bars and arrows 
indicate respectively  measurement uncertainties and  upper limits
 ($2\sigma$). In the middle column panels, the reference lines represent the
  mean (solid) and standard deviation (short-dashed) of the SFE computed using 
 all the detections in the whole sample (see Sect.~4.2). 
 In right column panels, the dashed  and continuous lines 
 show the $B_{1.4}$ and \ico trends, respectively.

\begin{figure}[t]
\begin{center}
\includegraphics[angle=0, width=9cm]{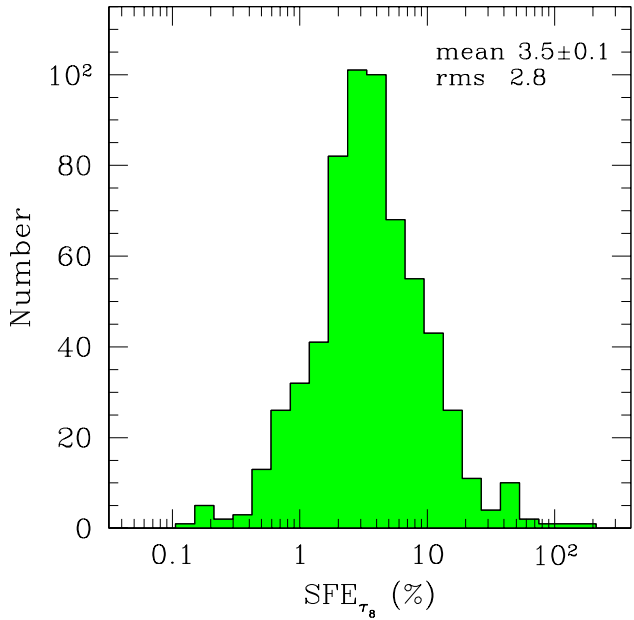}
\end{center}
\caption[]{Histogram of star formation efficiency at 628 pointing positions
for the entire sample. Only points with a signal-to-noise ratio greater
 than 2$\sigma$ have been considered. Mean and dispersion refer to the 
 logarithmic values of \sfe, see text. Values greater than
 100\% indicate present SFRs that cannot be supported for more than
 $10^{8}$ yr.}
\label{fig4}
\end{figure}
\subsection{The SFE within galaxy disks}
Most of the galaxies in our sample are at distances between 10 and 40 Mpc, 
the corresponding linear resolution of the 45\arcsec~beam is between 2.2 and
 8.7 kpc, i.e. so large to sample the contribution from
 many molecular clouds complexes. In the following, one should keep in mind
 that the star-forming regions we are observing are comparable with
 the width of the spiral arms only for the nearest objects. 
     
The galaxies shown in Figs. 2 and 3 are representative of the diversity of
 behaviour seen in the distributions of $B_{1.4}$ and \ico within galaxy disks.
  
The most striking feature is the linear correlation between these two 
quantities observed for many edge-on and face-on galaxies (see Fig.~2).
In these cases, $B_{1.4}$ and \ico present the same scaling from the galaxy
 center outward,  resulting in a constancy of the SFE along the disk.
However, there are clear examples of disks characterized by systematic SFE
 trends (see Fig.~3).
A \b14 gradient steeper (flatter) with respect to \ico one implies a SFE
 decreasing  (increasing) with radius (e.g. NGC5236 and NGC5247).
Calculating the SFR from non-thermal radio continuum allows us to include
 in the analysis high inclined galaxies, such as NGC1055 and NGC3079,
 which generally suffer from extinction in the optical band (see Sect.~5.1).
By fitting a power law of the form $B_{1.4}\propto {I_{\rm CO}}^{N}$, we found 
that the fraction of linear ($0.5< N <2$) correlations is 67\%,
while the fractions of super-linear ($N>2$) and sub-linear ($N<0.5$) correlations are 23\% and 10\%, respectively.  We examine the composite correlation 
including all the pointings in the sample in Sect.~4.3.
\begin{figure}[t]
\begin{center}
\includegraphics[angle=0, width=9cm]{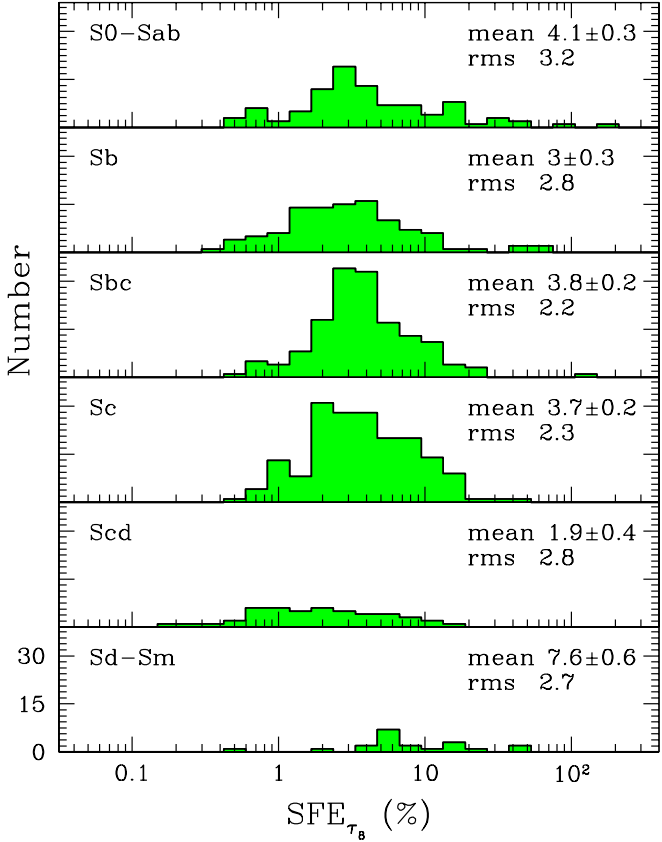}
\end{center}
\caption[]{Histograms of star formation efficiency sorted by galaxy type.
 Only points with a signal-to-noise ratio greater
 than 2$\sigma$ have been considered. Mean and dispersion refer to the 
 logarithmic values of \sfe, see text.}
\label{fig5}
\end{figure}	 
\subsection{The SFE variation among galaxies}
Having established a correlation between the radio continuum surface
 brightness and the CO line intensity within galaxies, we investigated
 the variation of the SFE among galaxies. Fig.~4 shows the
 distribution of the \sfe~for the galaxy sample.
\sfe varies from about 0.1 up to more than 100\%. Values greater than
 100\% indicate present SFRs that cannot be supported for more than
 $10^{8}$ yr.
 Since the distribution of logarithmic values is more Gaussian we calculated 
the mean and the dispersion of $\log$(\sfe). 
The mean of  $\log$(\sfe)~is approximately 3.5\% of the gas consumed
 per 10$^{8}$ yr and the rms of the distribution is slightly less than a factor
 of 3. \sfe~mean and dispersion of the sample are traced in middle plots 
of Figs. 2 and 3  as solid and dashed reference lines, respectively.   

 We examined also the variation of \sfe~among galaxies compared to 
the morphological type (see Fig.~5).The mean star formation efficiency varies
 weakly (about 25\%) with the morphological type going from
 S0 to Scd galaxies.
\begin{figure}[h]
\begin{center}
\includegraphics[angle=0, width=9cm]{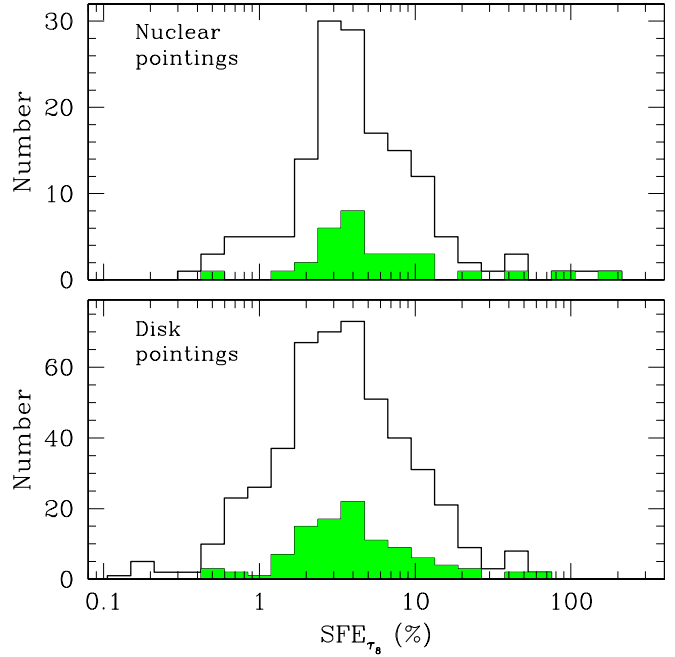}
\end{center}
\caption[]{Histograms of star formation efficiency for nuclear (top panel) and
 disk (bottom panel) pointings. The solid line and solid 
portion of the histograms indicate normal and Seyfert galaxies, respectively.
Three galaxies have nuclear \sfe~greater than 100\%;
 namely NGC2655, NGC3310 and NGC4151. NGC2655 and NGC4151 are classified as
Sy2 and Sy1.5 and have nuclear \sfe~of 100\% and 170\%, respectively. 
The galaxy NGC3310 is a peculiar starburst galaxy characterized by an
 exceptional nuclear \sfe~of about 135\%}
\label{fig6}
\end{figure}

Larger SFE close to the galaxy centers might be attributed to the
presence of an active nucleus (AGN). Fig.~6 shows the 
distribution of disk and nuclear \sfe~in Seyfert and normal galaxies.
Seyferts have a slightly higher \sfe~than normal galaxies.
The mean of the log \sfe, for Seyfert and normal galaxies, is: 4.1\% and 3.3\% in 
the disks and 5.5\% and 4.3\% in the nuclei, respectively.
Most Seyferts have a nuclear \sfe~comparable with normal galaxies. Only in 
few cases the nuclear emission is dominated by the radio source
related with the AGN, e.g. NGC1068 (Wilson \& Ulvestad 1987), NGC4151 (Pedlar
 et al. 1993), NGC2655 (Keel \& Hummel 1988) and NGC4388 (Irwin et al. 2000).
We conclude that the AGN-related emission affects only marginally the estimate 
of the \sfe~in the nuclear pointings for most Seyfert in our sample.

We further investigate the behaviour of the SFE with respect to the galaxy 
inclination and size, and beam linear resolution. Fig.~7 shows the maximum 
variation of the \sfe, defined as SFE$_{\rm max}$/SFE$_{\rm min}$,
 inside each galaxy. Most galaxies present SFE variation  
 up to a factor 6,  the median variation being a factor of 2.5. 
\begin{figure}[b]
\begin{center}
\includegraphics[angle=0, width=9cm]{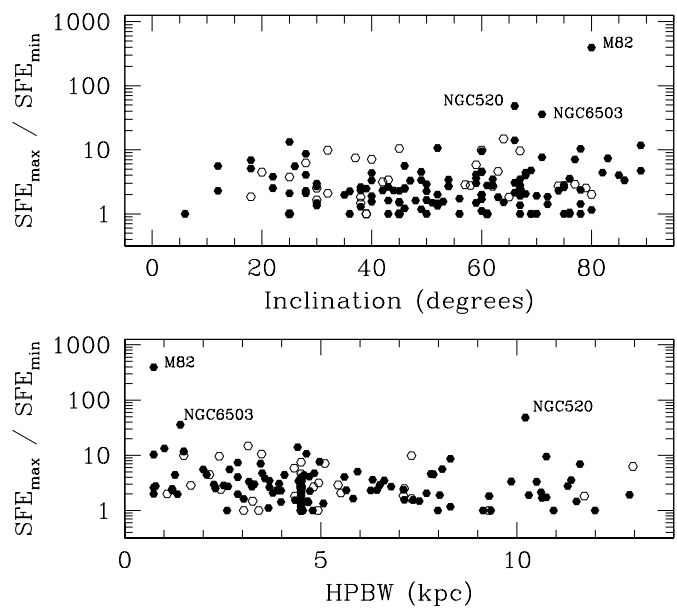}
\end{center}
\caption[]{Star formation efficiency variations within each 
 galaxy as a function of inclination (top panel) and linear resolution of
 the 45\arcsec~beam (bottom panel). Seyfert galaxies are distinguished with
 open dots. The majority of galaxies have internal SFE variation less than
 a factor of 3. The starburst galaxies NGC3034 (M82) and NGC520 and the 
HII galaxy NGC6503, have exceptional internal SFE variations 
(larger than a factor of 30).}
\label{fig7}
\end{figure}

Fig.~7 shows that the internal SFE variations are not strong function of
 the galaxy inclination or linear resolution of the observing beam. 
Considerable SFE variations occur also for size of the 
 beam greater than 5 kpc, i.e. smoothing over large regions of the galaxy disks.
 This fact was already noted by RY99. 
Ten galaxies show an internal SFE variation greater than a factor of 10. These
 are: the circumnuclear starbursts IC342, NGC253, NGC520, NGC660, NGC2146 and 
 NGC3034 (see Kennicutt 1998); the Seyfert galaxies NGC2841 and NGC3368;
 the peculiar galaxy NGC3628; the HII galaxy NGC6503. 
 M82, NGC520 and NGC6503, show exceptional internal SFE variations larger
 than a factor of 30. In particular, the nearby starburst M82 show a variation
 of about 2.6 order of magnitude. 

Finally, we investigated the behaviour of the star formation efficiency as
 a function of the distance from the galaxy centers.
Fig.~8 shows the SFE as a function of radius for all galaxies. 
The SFE is found to be approximately constant at all radii.

\begin{figure}[t]
\begin{center}
\includegraphics[angle=0, width=8.5cm]{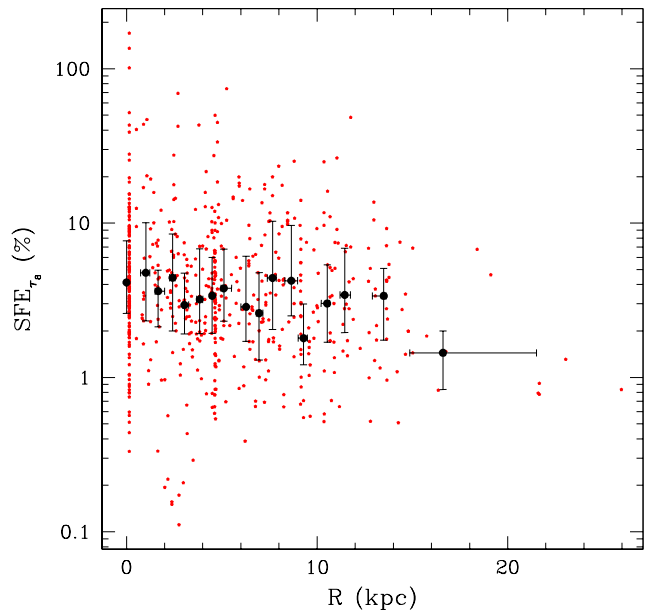}
\end{center}
\caption[]{Star formation efficiency as a function of radius for all
 galaxies. Limits are not shown. All the detections are represented
as small points.
Bold points and error bars represent the median value and the 50\% of 
objects in each bin, respectively.}

\label{fig8}
\end{figure}
\subsection{The composite radio Schmidt law}
Our data offer the possibility to determine the 
 {\it spatially resolved} Schmidt law using the radio continuum as
 SFR indicator for a sample of galaxies six times larger than the one of Adler et al. (1991).
 Fig.~9 shows
 the relationship between the SFR and the \hh surface densities for all
 the pointings of the sample. Dashed and dotted reference lines indicate
 the mean and standard deviation of the \sfe~for all the detection in the
 sample (see Sect.~4.1).
 The galaxies NGC4302, NGC4526, NGC4710 and
 NGC4866 have been omitted since they have inclination $i=90\degr$. A clear
 correlation is evident over more than three orders of magnitude, albeit a
 consistent scatter of 0.5 dex is present. We stress once more that, since 
this is a correlation between two brightness, we can rule out any subtle
 effect introduced by the distance. This result confirms the
 correlation found by Adler et al. (1991) on global scales. 
The best weighted fit of the radio Schmidt law, 
$\Sigma_{{\rm SFR}}=a \times {\Sigma_{{\rm H_{2}}}}^{N}$, 
is: $a=2.6(\pm1.2)\times 10^{-4}$ and $N=1.3\pm 0.1$.
 The value for the exponent of the radio Schmidt law is
 fully consistent with both the global (e.g. Kennicutt 1998) and local
 (RY99) slopes found using the \ha line emission as SFR 
indicator. This exponent for the Schmidt law indicates a SFE which weakly 
increases with the SFR.   

\begin{figure}[t]
\begin{center}
\includegraphics[angle=0, width=8.5cm]{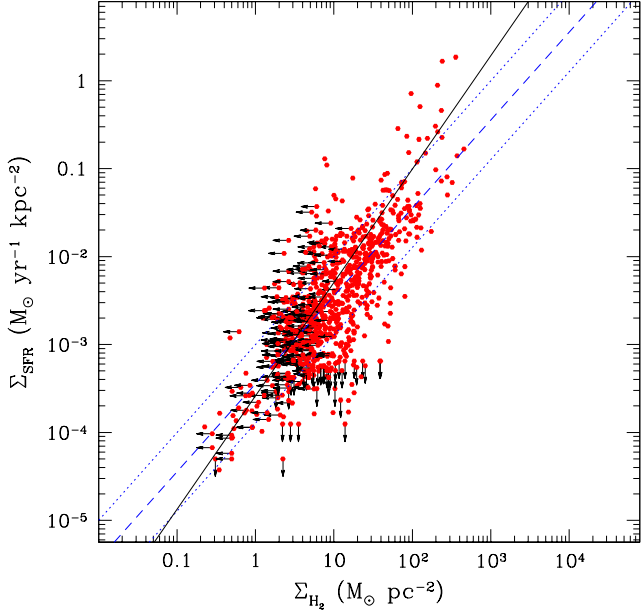}
\end{center}
\caption[]{Spatially resolved radio Schmidt law for all the detections
 in the sample. Arrows indicate upper limits at 2$\sigma$.
 The solid line is the best fit of the composite radio Schmidt law: 
$\Sigma_{{\rm SFR}} \propto {\Sigma_{{\rm H_{2}}}}^{1.3}$.
 Dashed and dotted lines refer to the mean and standard deviation of
 the \sfe~for all detections in the sample: 3.5\% and a factor of 2.8
 respectively.  
  The galaxies NGC4302, NGC4526, NGC4710 and NGC4866 
have been omitted since they have inclination $i=90\degr$.}
\label{fig9}
\end{figure}

\section{Discussion and conclusions}
The most striking result of our study    
is the overall consistency of the SFE obtained from the 20~cm radio continuum 
as SFR indicator instead of the \ha~emission (RY99).
Basically we found all the essential features observed by RY99: the constancy 
of the SFE both within and among galaxies; the weak dependence of the 
SFE on morphological type and linear size of the observing 
 beam; the extent and slope 
of the composite Schmidt law. Nevertheless, some important differences
 arise. In this section we present in details the differences
 between the two works, discussing their nature and implications.

\begin{figure}[t]
\begin{center}
\includegraphics[angle=0, width=8.5cm]{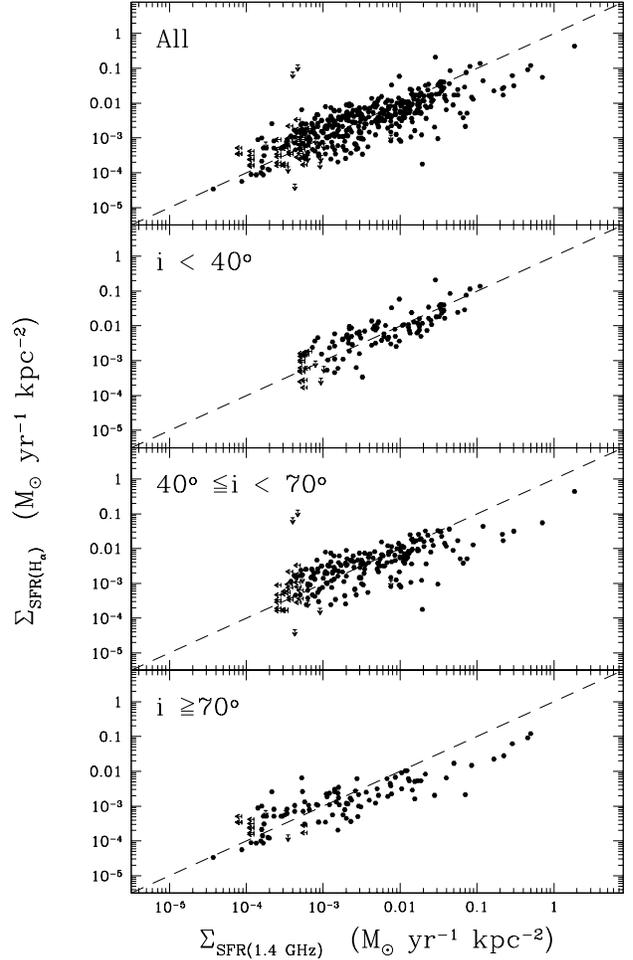}
\end{center}
\caption[]{
Comparison of SFRs surface densities deduced from 1.4 GHz luminosity
 (horizontal axis) and \ha emission (vertical axis) for 102 galaxies of
 our sample in common with RY99. 
The various panels show different values of galaxy
 inclinations $i$, as indicated. There are 453 detections, 65 of which are
 upper limits (at 2$\sigma$).
The reference dashed lines indicate 1:1 relations. 
The subsample of face-on galaxies ($i \leq 40\degr$) shows the best correlation,
 with a dispersion of a factor of 2.3.
SFRs surface densities deduced by the \ha for highly inclined galaxies
 ($i \geq 70\degr$) are systematically underestimated for 
$\Sigma_{\rm SFR} {\rm 1.4~GHz} \geq 3\times 10^{-3} ~{\rm M_{\odot}~yr
^{-1}kpc^{-2}}$.}
\label{fig10}
\end{figure}

\subsection{Star formation rate surface densities}
Cram et al. (1998) checked the consistency between the SFR deduced from the
 radio luminosity with the rates predicted by other indicators, in particular
 by the \ha luminosity. They concluded that the rates deduced from the radio 
continuum and the \ha luminosities, although in broad agreement, are affected
 by two systematic deviations: respectively at low 
($\leq 0.1 ~{\rm M_{\odot}yr^{-1}}$) and high ($\geq 20 ~{\rm M_{\odot}yr^{-1}}$)
 star formation rates, the \ha luminosity overestimate and
 underestimate the SFR deduced by the radio luminosity. 
The deviation at low rates can be attributed in part to problems 
 related to the zero-point \ha luminosity calibration.
They suggest that the deviation at high SFR could be attributed to
 a larger amount of extinction by dust in those objects undergoing
 strong star formation or to particular IMFs that favour low mass supernovae
 progenitors rather than high mass stars responsible for the \ha. 

By comparing our data set with that of RY99 we have the 
 possibility to extend the consistency check between the surface star
 formation rate densities deduced from the radio continuum and  \ha emission
 over the same regions of galaxy disks.

\begin{figure}[b]
\begin{center}
\includegraphics[angle=0, width=9cm]{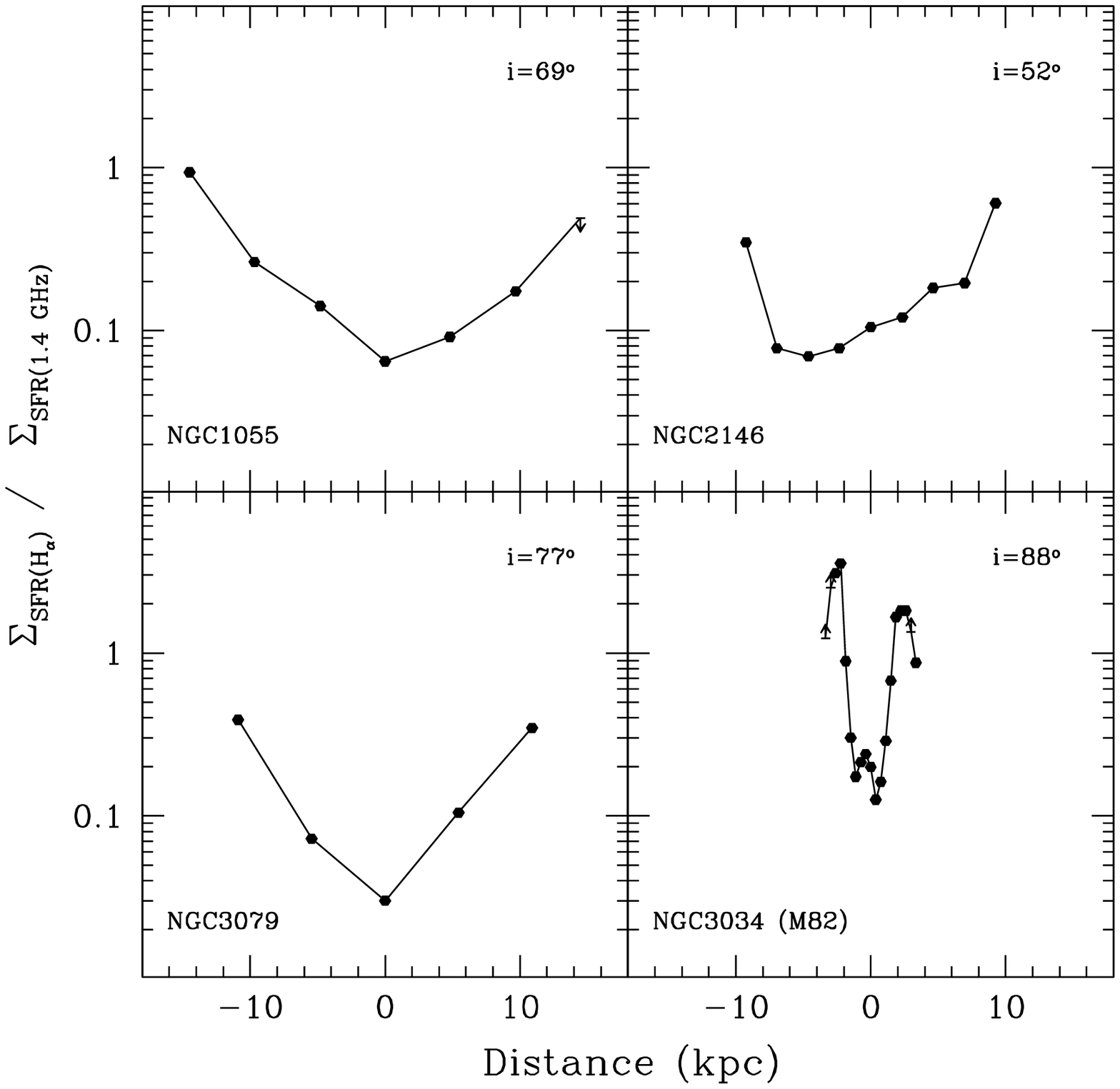}
\end{center}
\caption[]{
Ratio of SFRs surface densities deduced from \ha and 1.4 GHz luminosity 
 for  highly inclined (NGC1055 and NGC3079) and starburst
 galaxies (NGC2146, NGC3034).
 These examples dramatically illustrate that
 the SFRs surface density inferred from the
 \ha emission towards the nuclei of these objects are underestimated 
by more than one order of magnitude.}
\label{fig11}
\end{figure}
Using Eq.~(2) of Cram et al. (1998), we calculated the SFR surface density
 from the \ha brightness reported by RY99 through the
 formula:
\begin{equation}
\Sigma_{\rm SFR}\left({\rm \frac{M_{\odot}}{yr~kpc^{2}}}\right) =
2.6\times10^{-2}~\mu({\rm H_{\alpha}})
\label{hasfr}
\end{equation}
where the (de-projected) \ha surface brightness, $\mu({\rm H_{\alpha}})$,
 is measured in ${\rm L_{\odot}~pc^{-2}}$.
 Eq.~(2) of Cram et al. (1998) accounts for a 
correction of 1.1 magnitude in the luminosity to compensate for the 
mean extinction in \ha (Kennicutt 1983a).
Fig.~10 shows the direct comparison between the radio and \ha SFRs inferred
 respectively from Eq.~(\ref{sfrb}) and Eq.~(\ref{hasfr}) for 453 position
 within 102 galaxies of the sample. All galaxies are
 shown in the top panel of Fig.~10. In general there is a broad agreement 
between the two SFR indicators.
Considering the distribution of the logarithm values of the  ratio
 $R=\Sigma_{\rm SFR(H_{\alpha})}/\Sigma_{\rm SFR(1.4 GHz)}$, we have that 
 $R$ have a mean of 0.83 with a standard deviation of 2.8, i.e. 
the \ha is underestimating
 the surface SFR as compared to continuum radio emission. 
Sorting the sample by galaxy inclination indicates that this effect is 
 strongly correlated with galaxy orientation. For the face-on subsample 
($i < 40\degr$) $R$ has a mean of 1.1 and a dispersion of a factor 2.3. At 
intermediate inclinations ($40\degr \leq i < 70\degr$) mean 
 and dispersion of $R$ are 0.8 and a factor of 3,
 respectively. Finally,  for highly 
 inclined galaxies ($i \geq 70\degr$)  mean 
 and dispersion of $R$ are  0.7 and a factor of 2.9, respectively.
 In particular, SFR surface densities
 deduced by the \ha for highly inclined galaxies are
 systematically underestimated for $\Sigma_{\rm SFR(1.4~GHz)}$ greater 
 than  $\sim 3\times 10^{-3} ~{\rm M_{\odot}~yr^{-1}kpc^{-2}}$.
Fig.~11 shows two edge-on spirals, NGC3079 and NGC1055 (see Figs.
 2 and 3) and two well known starburst galaxies, NGC2146 and M82, 
 in which  SFRs deduced by the \ha emission are underestimated by more than
 a factor of about 10. As a consequence, the SFE variations
 deduced for these objects by RY99, on the basis of
 \ha observations, can be attributed to extinction, as suspected by these 
authors.

These results have two important implications: i) the close correlation  
  observed between $\Sigma_{\rm SFR(1.4~GHz)}$ and
 $\Sigma_{\rm SFR(H_{\alpha})}$ for face-on
 galaxies reinforces the use of the radio luminosity
 as SFR indicator not only on global but also on local scales; ii)
  extinction could significantly affect estimates
 based on \ha emission for high SFRs in edge-on galaxies.
 
Although the star formation rates deduced from the \ha emission are 
systematically underestimated compared to those deduced from the radio
 continuum, the mean SFE reported by RY99 for their entire sample 
 (121 galaxies) is 4.3\%, i.e. higher than the mean SFE deduced from the radio
 continuum for our entire sample of 180 galaxies. However, considering
 the 103 galaxies we have in common with RY99, the mean SFE deduced from 
the \ha emission is about 3.1\% in good agreement with our value.
 
\begin{figure}[h]
\begin{center}
\includegraphics[angle=0, width=9cm]{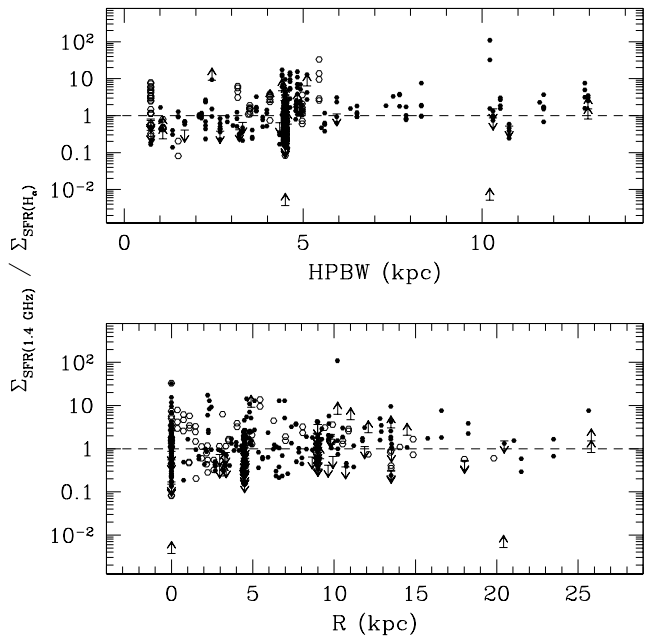}
\end{center}
\caption[]{
Ratio of SFRs surface densities deduced from 1.4 GHz luminosity and \ha
 emission as functions of the linear scale of the beam
 (top panel) and distance from galaxy center (bottom panel). Highly inclined
 galaxies ($i> 70\degr$) are represented with open dots.}
\label{fig12}
\end{figure}
\subsection{Non-thermal radio continuum scale-length}
Discrete supernova remnants (SNR) themselves account only for $<10\%$ of
 the radio luminosity of normal galaxies (Ilovaisky \& Lequeux 1972).
 Most of cosmic rays escape from their parent SNRs and diffuse along the 
galaxy disk during their characteristic lifetimes which are much larger than
 typical ages of SNRs ($\sim 10^{5}$ yr). Therefore one expects the 
non-thermal radio continuum emission to be smoothed over a larger region
 around the parent stellar population with respect to the \ha emission.
On scale smaller than the cosmic rays characteristic diffusion scale, $D_{\rm CR}$,
 the non-thermal radio continuum cannot provide a reliable estimate of
 the local SFR. On the other hand, averaging over regions larger than the
 typical spiral arm width leads to a dilution of the \ha emission.
Fig.~12 (top panel) shows the ratio between the SFR surface
 densities  deduced  from \ha and from radio luminosity versus
 linear size of the observing beam. One notes that, apart for edge-on galaxies
 (where the \ha is affected by extinction), the ratio $\Sigma_{\rm SFR(1.4~GHz)}$ to  $\Sigma_{\rm SFR(H_{\alpha})}$ tends to increase with
 the beam size. 
In particular, the ratio is smaller or greater than one for beam sizes
 respectively below and above $\sim 5$ kpc. This is consistent with a
 scenario in which the characteristic linear scale of both cosmic rays
 diffusion scale and arm size is about 3 kpc. Higher resolution radio images 
 are required to determine the lowest linear scale at which the 
radio continuum  correlates with the \ha emission and thus providing an
 estimate of $D_{\rm CR}$.

We found also that the ratio  $\Sigma_{\rm SFR(1.4~GHz)}$ to $\Sigma_{\rm SFR(H_{\alpha})}$ on average does not depend on the distance from
 the galaxy centers (Fig.~12, bottom panel).  

\subsection{The nature of Schmidt law scatter}
The dispersion (standard deviation) of the composite radio
 Schmidt law shown in Fig.~9 is a factor of 2.8, which is smaller than
 the corresponding dispersion of a factor 4 of the \ha Schmidt law (RY99). 
This is probably because the radio continuum does not suffer the  
effects of dust
 extinction. The noisiness of the correlation could reflect real variations
 of the mean Schmidt law, indicating that it should be regarded at best 
 as a first  order approximation of the star formation process, or it could
 be due to the many assumptions involved in the calculation of gas and SFR
 densities.

 Variations of the $X_{\rm CO}$ factor from galaxy to galaxy can be 
advocated to explain a part of the correlation scatter. These could introduce 
 uncertainties up to a factor of 2 in the gas density scale (Kennicutt 1998).
Another possibility is that the extrapolation of the proportionality 
between CO luminosity and virial mass of giant molecular clouds observed 
in our own Galaxy and in nearby galaxies, which is the basis of
 molecular mass determinations in this and similar works, does not hold
 for all spiral galaxies.
\begin{figure}[t]
\begin{center}
\includegraphics[angle=0, width=8.75cm]{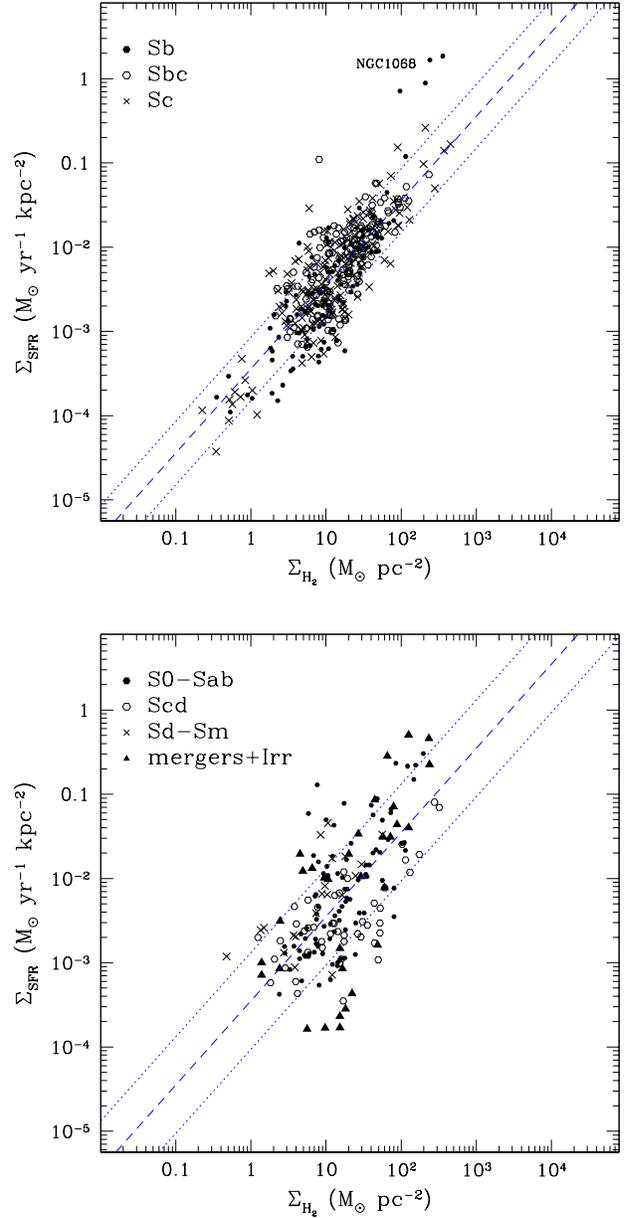}
\end{center}
\caption[]{
Radio Schmidt law for Sa, Sbc and Sc (top panel) and S0-Sab, Scd, Sd-Sm and
 merging-irregular (bottom panel); limits are not shown.
 Dashed and dotted reference lines indicate the mean and
  the dispersion of the two sub-samples, respectively. The scatter
 of the Schmidt law
 is considerably reduced excluding extreme morphological types.
 The scatter of the correlations shown in top and bottom panels is a factor 
of 2.4 and 3.8, respectively. In top panel, the deviation from the 
correlation of the four points belonging to NGC1068 is due to the AGN-related
 radio emission of this Seyfert galaxy (see Sect.~4.2).}  
\label{fig13}
\end{figure}

However, by comparing SFR surface densities deduced from \ha 
and from radio continuum luminosity we showed that, even in the face-on
 subsample, the $\Sigma_{\rm SFR(H_{\alpha})}-\Sigma_{\rm SFR(1.4~GHz)}$ 
relation itself is affected by a scatter of at least a factor of 2. Hence,
 the uncertainties of the SFR indicators alone can account for a consistent
 fraction of the Schmidt law scatter.

Fig.~13 shows the Schmidt law separately for Sb, Sbc and Sc and  S0-Sab, 
Scd and Sd-Sm galaxies
 (only detections with signal-to-noise ratio greater than 2$\sigma$ are
 considered). The two subsamples have the same mean SFE of 3.5\% but very 
different dispersions of a factor of 2.4 for the former and 3.8 for the latter,
 i.e. the scatter of the Schmidt law is considerably reduced excluding
 extreme morphological types. RY99 also found that the \ha Schmidt
 law is considerably tightened with the exclusion of the  
 irregular galaxies and mergers. These SFE variations around the mean Schmidt
 law can be attributed to the particular physical conditions and/or
 environments experienced by these objects (see e.g. starburst galaxies),
 but they could also be
 induced by observational effects further amplified by poor statistic.
In fact, Scd galaxies which are characterized by the lower mean
 SFE in our sample (see also Fig.~5) behave consistently  with other 
morphological types excluding IC342 for which the NVSS misses flux density
 from the extended structure, see Sec.~2.

\subsection{Molecular gas consumption timescale}
The star formation efficiency can be expressed in terms 
 of gas consumption (or cycling) timescale. The molecular gas consumption
 timescale, $\tau_{\rm H_{2}}$, indicates the time needed to convert all
 the molecular gas in stars given a constant SFE. Expressed in Gyr, 
$\tau_{\rm H_{2}}$ is given by
\begin{equation}
\tau_{\rm H_{2}}({\rm Gyr})=\frac{10}{\rm SFE_{\tau 8} (\%)}
\end{equation}
The mean star formation efficiency found in this work for spiral galaxies  
yields a molecular gas consumption timescale of about $\tau_{\rm H_{2}}
\simeq 2.8$ Gyr. Given the dispersion of the SFE in the sample,
 most galaxies have gas cycling timescale between about 1 and 9 Gyr.
 At the lower end of  the distribution we found galaxies with a 
 \sfe$\sim 1\%$, such as NGC4535 (see Fig.~2).
 The corresponding molecular gas consumption timescale is
 almost comparable with the Hubble time. The shortest gas cycling timescale 
 corresponds to galaxies with exceptional SFE like NGC3310 (see Fig.~3). 
 Star formation rates in these galaxies are so high, compared to their gas
 supplies, that the gas cycling timescale is  $\tau_{\rm H_{2}} \le 0.1$
 Gyr.

It is interesting to investigate the behaviour of the so-called starburst
 galaxies with respect to the gas cycling timescale. In literature,
 galaxies have been classified as starbursts according to different criteria.
 Heckman et al. (1998) define a galaxy as starburst when it is hosting
 a star-forming event that dominates its bolometric luminosity, i.e. on the
 basis of the magnitude of the SFR. Alternatively, Shu (1987) and Young (1987) 
proposed a classification based on the efficiency of the star formation. In
 this latter definition a galaxy with a high SFR is not defined as starburst 
 if the mass of gas available is enough to sustain the star formation rate and 
vice versa. Following RY99 we selected the galaxies 
 hosting a region in which the SFE is enhanced by a factor of
 three compared 
 to the mean of the sample; for these starburst regions 
$\tau_{\rm H_{2}} \le 1$ Gyr. Fig.~14 shows the usual $\Sigma_{{\rm SFR}} - 
\Sigma_{{\rm H_{2}}}$ plane for all the detection in the sample along with three 
reference lines indicating the gas cycling timescales
 $\tau_{\rm H_{2}}=0.1$, 1 and 10 Gyr. 
\begin{figure}[t]
\begin{center}
\includegraphics[angle=0, width=9cm]{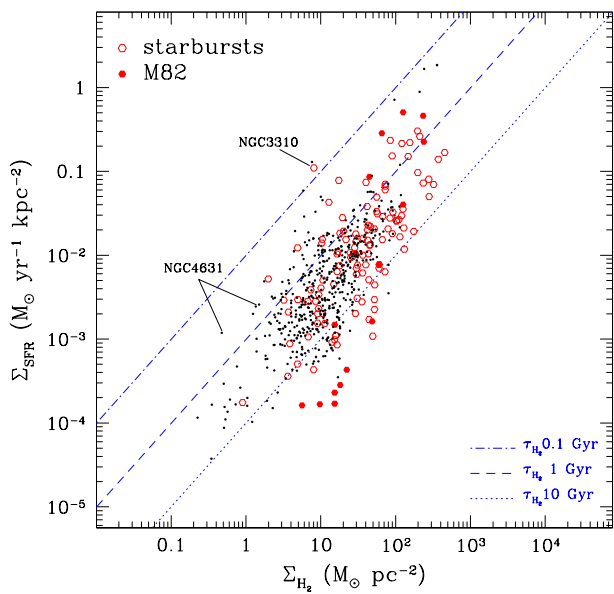}
\end{center}
\caption[]{
$\Sigma_{{\rm SFR}} - \Sigma_{{\rm H_{2}}}$ plane for all the detection in the
 sample (limits are not shown) along with three reference lines 
corresponding to different molecular gas consumption timescales, as indicated.
Pointings of known starburst galaxies are represented by open dots.
}
\label{fig14}
\end{figure}
Most galaxies known in literature as ``starbursts'' have consumption
 timescales comparable with those of normal spiral galaxies. In some cases,
 e.g. M82, starburst galaxies host both regions characterized by SFE 
 lower and higher than the mean of the sample. On the other hand, 
galaxies with normal or even low SFR surface densities, such as the highly 
inclined galaxy NGC4631, should be regarded as starburst according to 
our selection criterion based on the SFE.
Finally, there are some galaxies for which the SFE is so high that 
 the gas cycling timescale is $\tau_{\rm H_{2}} \le 0.1$
 Gyr, e.g. the starburst NGC3310. NGC3310 is one of the best examples 
of a local UV-bright starburst (see Conselice et al. 2000 and reference
 therein). The exceptional high SFE of this galaxy indicates that  
its interstellar medium is truly affected by an extraordinary star
 formation event. 

\subsection{Conclusions}
The results of this work are the following:

1.~There is a tight correlation between the 20 cm non-thermal
 radio continuum and the CO line intensity in a representative sample of 180 spiral galaxies. The correlation holds within and among the galaxies.

2.~The mean star formation efficiency, i.e. the ratio of the
 radio SFR to the molecular gas densities,  for our sample is $0.035\times
 10^{-8}$ yr$^{-1}$ with a dispersion of a factor of 3. This corresponds
 to convert 3.5\% of the available gas into stars on a time of $10^{8}$ yr.

3.~The comparison of SFRs surface densities deduced from 1.4 GHz 
luminosity and from the \ha emission for 102 galaxies, reveals that
  $\Sigma_{\rm SFR(1.4~GHz)}$ and $\Sigma_{\rm SFR(H_{\alpha})}$ are
 closely correlated for face-on galaxies ($i \leq 40\degr$), reinforcing
the use of the radio radio luminosity as SFR indicator not only on
global but also on local scales.
 SFRs surface densities deduced by the \ha luminosity 
for highly inclined galaxies ($i \geq 70\degr$) are systematically 
underestimated for
 $\Sigma_{\rm SFR(1.4~GHz)} \geq 3\times 10^{-3} ~{\rm M_{\odot}~yr
^{-1}kpc^{-2}}$.

4.~The  star formation efficiency varies weakly (less than 25\%)
 with the Hubble morphological type. 

5.~The variation of the SFE within individual galaxy disks is less
 than a factor of 3. The largest variations are found in starburst galaxies.

6.~The SFE is found to be approximately constant as 
a function of distance from the galaxy centers.

7.~The composite radio Schmidt law, star formation versus molecular
 gas content,  extends for more than 3 order of magnitude with an exponent
 of 1.3.

8.~Most galaxies known in literature 
as ``starbursts'' have consumption timescales comparable with those of normal
 spiral galaxies. In some cases, e.g. M82, starburst galaxies host both 
regions characterized by a SFE lower and higher than the mean of the
 sample. Furthermore, there are some galaxies for which the SFE is so high 
that the gas cycling timescale is $\tau_{\rm H_{2}} \le 0.1$ Gyr, e.g. 
NGC3310.

\begin{acknowledgements}
We thank R. Fanti, G. Grueff and M. Johnson who carefully read the 
manuscript and provided useful comments.
We acknowledge the Italian Ministry for
University and Scientific Research (MURST) for partial financial support
(grant Cofin99-02-37).
The National Radio Astronomy Observatory is operated by Associated 
Universities, Inc., under contract with National Science Foundation.
\end{acknowledgements}

\end{document}